# Micromanipulation of magnetotactic bacteria with a microelectromagnet matrix

H. Lee, A. M. Purdon, V. Chu, and R. M. Westervelt[a]

*Division of Engineering and Applied Sciences and Department of Physics,*

*Harvard University, Cambridge, Massachusetts 02138*

**Abstract**

Micromanipulation of magnetotactic bacteria with a microelectromagnet matrix was demonstrated. Magnetotactic bacteria synthesize a chain of magnetic nanoparticles inside their body to guide their motion in the geomagnetic field. A microelectromagnet matrix consists of two arrays of lithographically patterned wires, one array perpendicular to the other, that are separated and covered by insulating layers. By adjusting the current in each wire, a matrix can create versatile magnetic field patterns on microscopic length scale. Using a matrix, magnetotactic bacteria were trapped, continuously moved, rotated, and assembled in water at room temperature.

---

[a] Electronic mail: westervelt@deas.harvard.edu

With the advent of the controlled synthesis of nanocrystals[1], efforts are being made to manipulate these particles on microscopic length scales to build custom-made structures[2]. By trapping metallic[3] or semiconducting[4] nanoparticles between electrodes, single-electron devices were constructed. Genetically engineered viruses were used to assemble semiconducting nanocrystals into ordered structures[5]. Manipulation of magnetic nanoparticles is also of significant interest, because single-domain magnets have applications in spintronics[6], magnetic memory[7], and biology[8]. Using either permanent magnets[9] or electromagnets[10], magnetic manipulation tools have been implemented to control the motion of magnetic particles in a fluid. These tools were based on macroscopic external magnets, which limited the variety of magnetic field patterns that could be created.

A microelectromagnet matrix[11] was developed to micromanipulate magnetic particles in a fluid. A matrix consists of two layers of straight wires aligned perpendicular to each other, separated and capped by insulating layers. By controlling currents in the wires, a matrix can create versatile magnetic field patterns on microscopic length scales to trap, continuously move, and assemble magnetic particles suspended in a fluid. A matrix was used to micromanipulate magnetic beads in a fluid at room temperature[11].

In this letter, we show how the motion of magnetotactic bacteria can be controlled by a microelectromagnet matrix in water at room temperature. Through highly controlled biomineralization, magnetotactic bacteria[12] synthesize chains of intracellular, single-domain magnetic nanoparticles with permanent magnetic moments[13]. These chains allow the bacteria to passively orient themselves along the geomagnetic field line. Here, we demonstrate the micromanipulation of the bacteria using a matrix: a single group of bacteria was trapped and moved continuously, multiple groups of bacteria were moved independently, and a single bacterium was rotated. Using the bacteria as a building block, it could be possible to construct custom-designed magnetic structures from the biogenic magnetic nanoparticles contained inside magnetotactic bacteria with a matrix.

Figure 1(a) shows a micrograph of a microelectromagnet matrix fabricated on Si/SiO$_2$ substrate as described in Ref. 11. The matrix has two arrays of straight conducting wires, aligned perpendicular to each other as shown in the close-up, Fig. 1(b). Conducting wires were patterned using either optical lithography or electron beam lithography followed by Cr/Au deposition and lift-off. The width and the pitch of the wires can be adjusted, depending on the size of the object to be manipulated. As shown in the cross-sectional micrograph Fig. 1(c), each conducting wire layer is capped with an insulating layer, which prevents electrical shorting between wires. To reduce the friction between the magnetic objects and the surface of the device, a resin with a good planarization property (bisbenzocyclobutene) was used to form the insulating layers. The final device was treated with an O$_2$ plasma to make the surface hydrophilic, and a fluidic chamber fabricated with polydimethylsiloxanes using soft lithography[14] was attached on top.

With currents in the wires, microelectromagnets generate a local magnetic field pattern that can trap magnetic objects suspended in a fluid on the surface of the device. The magnitude of the magnetic field above a wire is $B \propto I/d$, where $I$ is the wire current and $d$ is the distance between the wire and the surface of the device. For the microelectromagnet reported here, the magnetic field magnitudes up to $B \sim 0.1$ T were achieved with current densities as high as $10^7$ A/cm$^2$ in the wires. The device was cooled by a thermoelectric cooler to prevent thermal breakdown from Joule heating, and electromigration. The potential energy of a trapped object is $U = -mB$, where $m$ is the magnetic moment of the object. In thermal equilibrium, the chemical potentials of magnetic objects inside and outside of a magnetic trap are equal to each other, giving the number density of objects inside a trap $n_T = n_O\exp(|U|/k_BT)$, where $k_B$ is the Boltzmann constant, $T$ is the temperature, and $n_O$ is the number density of objects outside the trap. In a magnetic trap with trapping volume $V_T$, one or more objects will be trapped provided $n_OV_T \geq \exp(-|U|/k_BT)$, which leads to minimum magnetic moment $m \geq -(k_BT/B)\log n_OV_T$ required for stably trapping. Using the experimental conditions reported here, $n_O = 10^{-7}/\mu m^3$, $V_T = 25$ $\mu m^3$ and $T = 288$ K, the minimum magnetic moment for stable trapping is $m = 5\times10^{-19}$ Am$^2$ with $B = 0.1$ T.

Once a magnetic object is trapped, its motion can be controlled above the surface of the matrix by changing the magnetic field patterns. By individually controlling the current in each wire, a matrix can generate versatile magnetic field patterns, allowing dynamic and complex manipulations of trapped objects. Figure 2 shows examples of magnetic field profiles calculated for a matrix. To produce a desired magnetic profile, the current in every wire of the matrix was optimized using least-square fits. In Fig. 2(a), a single peak in magnetic field magnitude, fit to a Gaussian profile, was created and moved continuously over the surface of the device. This profile can be used to trap and precisely position a magnetic object at a desired location. A matrix can also create multiple peaks simultaneously and control them independently. For example, in Fig. 2(b), three separate peaks were created and brought together at one position, while in Fig. 2(c) a single peak was split into four peaks that are moved apart independently. Using these magnetic field profiles, magnetic objects can be brought together to force interactions or sorted out to desired locations. Because each wire in a matrix can have a different current, optimized magnetic field patterns can be created to meet specific experimental needs without any modification in the structure of the device.

The micromanipulation of magnetotactic bacteria was performed using *Magnetospirillum magnetotacticum* (MS–1), a variety of magnetotactic bacteria that has a single chain of intracellular magnetite ($Fe_3O_4$) nanoparticles as shown in Fig. 3(a). The magnetic particles have a narrow size distribution with diameter ≈ 50 nm and a uniform shape as shown in Fig. 3(b), demonstrating highly controlled biomineralization by the species. In addition, the size of the magnetic particles falls in a range where each particle is a single-domain permanent magnet with magnetic moment $m_0 \sim 6 \times 10^{-17}$ $Am^2$. Adding up the individual magnetic moments $m_0$ of particles in the chain, the total magnetic moment of a bacterium is ~ $1.0 \times 10^{-15}$ $Am^2$, well above the minimum for stable micromanipulation in a fluid by a microelectromagnet matrix.

To image manipulation processes, the bacteria were stained with a green fluorescent dye and observed using a fluorescent microscope. A solution containing the stained bacteria was introduced into a fluidic chamber placed on top of the microelectromagnet. The temperature of the device was maintained at $T = 288$ K by a thermoelectric cooler attached on the back. To test the experimental setup, a simple ring trap (a circular wire covered with an insulating layer) was used to trap a single bacterium as shown in Fig. 3(c). Inside the trap, the bacterium underwent complex motions due to its own motility, but it remained trapped as long as the magnetic field ($B = 6$ mT at the center of the ring) was on. This operation demonstrates the noninvasive and stable trapping capability of microelectromagnets.

Figure 4 shows control of the motion of magnetotactic bacteria in water using a matrix with 10 wires in each layer (a 10×10 matrix). Currents were supplied by 20 current sources, one for each wire, that were individually controlled by a computer. In Fig. 4(a), a group of bacteria was trapped and then moved continuously over the surface of the device. Because the movements of the peak in magnetic field magnitude were continuous as shown in Fig. 2(a), the trapped bacteria could be moved over distances smaller than the wire spacing. Multiple groups of bacteria were controlled simultaneously as shown in Fig. 4(b). After initial trapping, a single group of bacteria was separated into two groups horizontally and then separated into four groups vertically. With time-varying currents, a matrix can generate dynamic field patterns as well. As an example, Fig. 4(c) shows the rotation of a single bacterium above the surface of the device. Two sinusoidal currents at the same frequency $f$ but with phase difference 90° were applied to two wires crossing each other. Trapping magnetic field was created at the crossing point of the two wires with its direction rotating at the same frequency. The sequence of images in Fig. 4(c) shows the rotation of a bacterium at $f = 0.1$ Hz.

These images demonstrate how a microelectromagnet matrix can be used to build custom-designed structures using magnetotactic bacteria. Magnetic nanoparticles with permanent magnetic moments are grown by the bacteria. The bacterial bodies enclose the nanomagnets and prevent clustering from magnetic dipole interactions, allowing controlled manipulation of a single chain of nanomagnets within a bacterium. After positioning and assembling the bacteria using a matrix, the cellular body of the bacteria can be removed by cell lysis, leaving the intracellular magnetic particles at desired locations. Combining biomineralization and micromanipulation, this new approach can open a way to grow and assemble magnetic nanoparticles into customized structures.

The authors thank X. Zhuang and M. Bawendi for their helpful comments. This work was supported by the Nanoscale Science and Engineering Center at Harvard under NSF grant PHY-0117795.

**Figures**

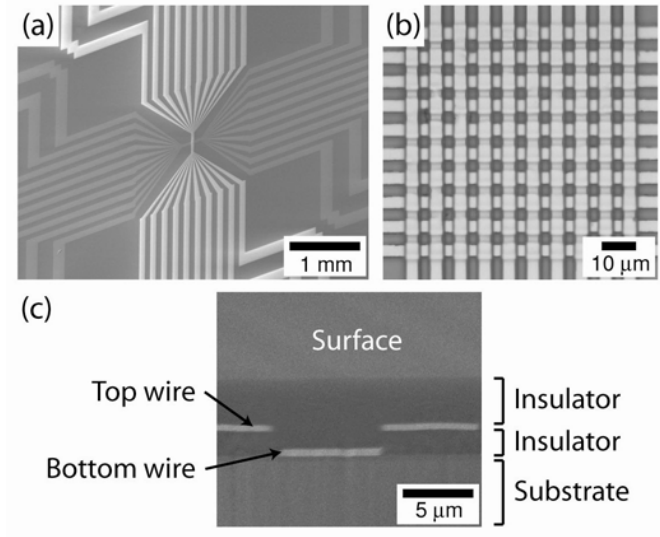

FIG. 1. (a) Micrograph of a 10×10 microelectromagnet matrix with its leads. (b) A close-up of the matrix: two layers, each with 10 Au wires, are aligned perpendicular to each other, separated and topped by insulating layers. The width and the pitch of the wires are 5 μm and 10 μm, respectively. (c) Cross section of the matrix along a diagonal direction; the top and bottom wires are indicated along with the substrate, two insulating layers, and the top surface.

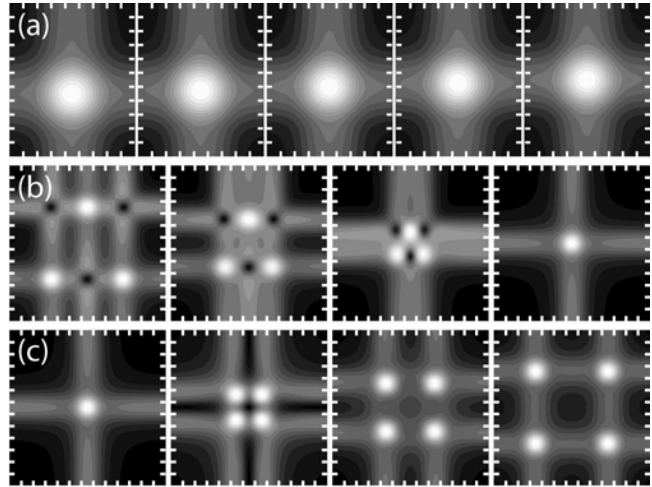

FIG. 2. Computed patterns of the magnitude of the magnetic field on the top surface of a matrix that demonstrate its principle of operation. The white ticks indicate wire positions and the total thickness of the two insulating layers is equal to the wire pitch. To produce a given magnetic pattern, the current in every wire was optimized using least-square fits. (a) A single peak in the magnetic field magnitude moves continuously between two adjacent wires. (b) Three separate peaks are brought together to converge at one position. (c) A single peak is split into four peaks that are moved apart independently.

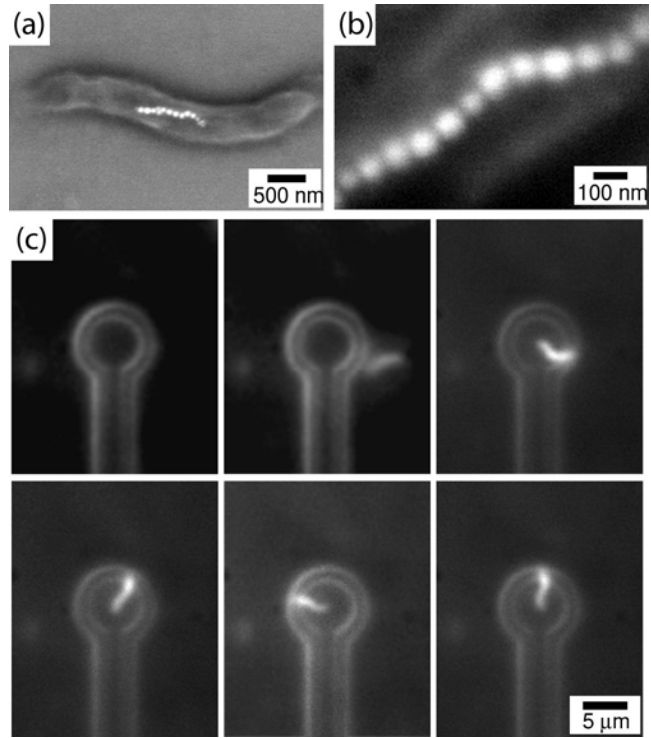

FIG. 3. (a) Micrograph of *Magnetospirillum magnetotacticum* (MS–1). Each bacterium synthesizes a chain of magnetite nanoparticles inside its body. (b) Close-up of the magnetite nanoparticles. Each particle is covered with a membrane and has magnetic moment ~ $6\times10^{-17}$ Am$^2$. The total magnetic moment of a bacterium is ~ $1.0\times10^{-15}$ Am$^2$. (c) Sequence of micrographs demonstrating a single MS–1 bacterium trapping with a ring trap. With $I = 50$ mA, a magnetic field peak $B = 6$ mT was created at the center of the ring. A single bacterium suspended in water was trapped and remained inside the trap with the magnetic field on.

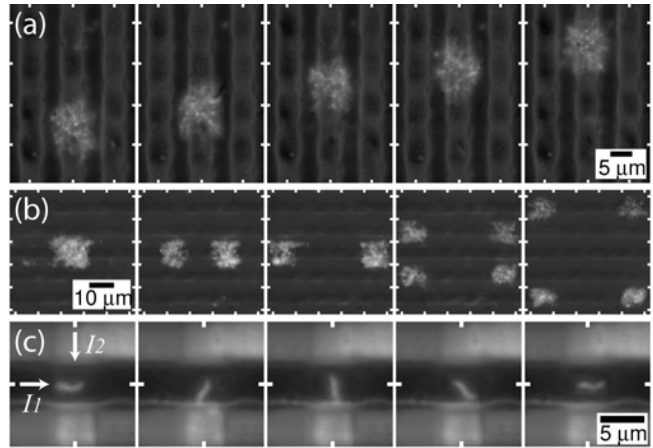

FIG. 4. Demonstration of the micromanipulation of MS–1 bacteria with a 10×10 wire microelectromagnet matrix. White ticks indicate wire positions. (a) A single magnetic field peak was created and moved continuously over two wires of the matrix, transporting a group of magnetotactic bacteria by increments smaller than the wire spacing. (b) A group of MS–1 bacteria was trapped, separated into two separate groups, and then further separated into four groups. In (a) and (b), currents in all 20 wires were adjusted to generate the desired magnetic field patterns. (c) Rotation of a single MS–1 bacterium at rotational frequency 0.1 Hz by two time-varying currents, $I_1$ and $I_2$ that have a 90° phase difference.

# References

1. V.I. Klimov, A.A. Mikhailovsky, S. Xu, A. Malko, J.A. Hollingsworth, C.A. Leatherdale, H.-J. Eisler, and M.G. Bawendi, Science **290**, 314 (2000).

2. S. Malynych, H. Robuck, and G. Chumanov, Nano Lett. **1**, 647 (2001); M.G. Warner and J.E. Hutchison, Nat. Mater. **2**, 272 (2003).

3. C.S. Wu, C.D. Chen, S.M. Shih, and W.F. Su, Appl. Phys. Lett. **81**, 4595 (2002).

4. D.L. Klein, R. Roth, A.K.L. Lim, A.P. Alivisatos, and P.L. McEuen, Nature **389**, 699 (1997).

5. S. W. Lee, C. Mao, C. E. Flynn, and A. M. Belcher, Science **296**, 892 (2002).

6. C.T. Black, C.B. Murray, R.L. Sandstrom, and S. Sun, Science **290**, 1131 (2000).

7. S. Sun, C.B. Murray, D. Weller, L. Folks, and A. Moser, Science **287**, 1989 (2000).

8. Urs Häfeli, *Scientific and clinical applications of magnetic carriers*. (Plenum Press, New York, 1997).

9. C. Gosse and V. Croquette, Biophys. J. **82**, 3314 (2002).

10. C. Haber and D. Wirtz, Rev. Sci. Instrum. **71**, 4561 (2000).

11. C.S. Lee, H. Lee, and R.M. Westervelt, Appl. Phys. Lett. **79**, 3308 (2001).

12. R. Blakemore, Science **190**, 377 (1975).

13. D. L. Balkwill, D. Maratea, and R. P. Blakemore, J. Bacteriol. **141**, 1399 (1980).

14. G. M. Whitesides, E. Ostuni, S. Takayama, X. Jiang, and D. E. Ingber, Annu. Rev. Biomed. Eng. **3**, 335 (2001).